\documentclass[conference]{IEEEtran}
\usepackage{cite}
\usepackage{amsmath,amssymb,amsfonts}
\usepackage{array}
\usepackage[normal]{caption}
\usepackage{seqsplit}
\usepackage{threeparttable}
\newcolumntype{R}[1]{>{\raggedleft\arraybackslash\hspace{0pt}\seqsplit}m{#1}}
\usepackage[table]{xcolor}
\usepackage{algorithmic}
\usepackage{graphicx}
\usepackage{textcomp}
\usepackage{quantikz}
\usepackage{xcolor}
\usepackage{flushend}

\newcommand\copyrighttext{%
  \footnotesize This work has been submitted to the IEEE for possible publication. Copyright may be transferred without notice, after which this version may no longer be accessible.} 
\newcommand\copyrightnotice{%
\begin{tikzpicture}[remember picture,overlay]
\node[anchor=south,yshift=10pt] at (current page.south) {\fbox{\parbox{\dimexpr\textwidth-\fboxsep-\fboxrule\relax}{\copyrighttext}}};
\end{tikzpicture}%
}

\begin{document}
\copyrightnotice

\title{Molecular Symmetry in VQE: A Dual Approach for Trapped-Ion Simulations of Benzene}

\author{\IEEEauthorblockN{1\textsuperscript{st} Joshua Goings}
\IEEEauthorblockA{\textit{IonQ, Inc.} \\
United States \\
goings@ionq.co}
\and
\IEEEauthorblockN{2\textsuperscript{nd} Luning Zhao}
\IEEEauthorblockA{\textit{IonQ, Inc.} \\
United States \\
zhao@ionq.co}
\and
\IEEEauthorblockN{3\textsuperscript{rd} Jacek Jakowski}
\IEEEauthorblockA{\textit{Oak Ridge National Laboratory} \\
United States \\
jakowskij@ornl.gov}
\and
\IEEEauthorblockN{4\textsuperscript{th} Titus Morris}
\IEEEauthorblockA{\textit{Oak Ridge National Laboratory} \\
United States \\
morristd@ornl.gov}
\and
\IEEEauthorblockN{5\textsuperscript{th} Raphael Pooser}
\IEEEauthorblockA{\textit{Oak Ridge National Laboratory} \\
United States \\
pooserrc@ornl.gov}
}

\maketitle

\begin{abstract}
Understanding complex chemical systems—such as biomolecules, catalysts, and novel materials—is a central goal of quantum simulations. Near-term strategies hinge on the use of variational quantum eigensolver (VQE) algorithms combined with a suitable ansatz. However, straightforward application of many chemically-inspired ansatze yields prohibitively deep circuits. In this work, we employ several circuit optimization methods tailored for trapped-ion quantum devices to enhance the feasibility of intricate chemical simulations. The techniques aim to lessen the depth of the unitary coupled cluster with singles and doubles (uCCSD) ansatz’s circuit compilation, a considerable challenge on current noisy quantum devices. Furthermore, we use symmetry-inspired classical post-selection methods to further refine the outcomes and minimize errors in energy measurements, without adding quantum overhead. Our strategies encompass optimal mapping from orbital to qubit, term reordering to minimize entangling gates, and the exploitation of molecular spin and point group symmetry to eliminate redundant parameters. The inclusion of error mitigation via post-selection based on known molecular symmetries improves the results to near milli-Hartree accuracy. These methods, when applied to a benzene molecule simulation, enabled the construction of an 8-qubit circuit with 69 two-qubit entangling operations, pushing the limits for variational quantum eigensolver (VQE) circuits executed on quantum hardware to date.
\footnote{ This manuscript has been authored in part by UT-Battelle, LLC, under contract DE-AC05-00OR22725 with the US Department of Energy (DOE). The publisher acknowledges the US government license to provide public access under the DOE Public Access Plan (https://energy.gov/doe-public-access-plan). }
\end{abstract}

\begin{IEEEkeywords}
Quantum Simulation, Variational Quantum Eigensolver, Unitary Coupled Cluster, Trapped-Ion Quantum Computer
\end{IEEEkeywords}

\section{Introduction}
Quantum computing's potential to provide more efficient solutions to complex electronic structure problems has drawn considerable attention.\cite{Nielsen2010-dm} The inherent challenge of these problems, which are ubiquitous across multiple industries from pharmaceuticals to material design,\cite{Boyd2007-sh,Poree2017-ih, Lam2020-vy, Blunt2022-la, Zinner2022-xh} lies in their unfavorable scaling with system size on classical resources. As the system size grows, conventional methods rapidly become impracticable due to their computational demand. Hence, there is a pressing need to explore scalable alternatives, with quantum computing presenting a compelling solution.

Specifically, quantum algorithms such as the Variational Quantum Eigensolver (VQE) have shown promising results.\cite{Peruzzo2014-xz, OMalley2016-ph, Colless2018-hi, McCaskey2019-gk, Nam2020-ct, Kandala2017-kr, Kandala2019-rz, Gao2019-pp, Rice2021-mn, Gao2021-zt, Zhao2023-um} VQE stands out due to its design for noisy intermediate scale quantum (NISQ) computers, employing shallow parameterized circuits that can be optimized classically to minimize energy expectation values. This makes VQE well-suited for tackling the current limitations of quantum hardware.

One cornerstone for the effective implementation of VQE is the choice of the ansatz. The unitary Coupled Cluster with singles and doubles (uCCSD) ansatz offers a recognized, capable method for treating strongly correlated systems,\cite{Anand2022-im} a prevalent case in electronic structure problems. In essence, it is a chemically-inspired approximation where the complexities of electron correlation are distilled into pair-wise interactions (doubles), along with individual electron promotions (singles). In the quantum realm, we translate this into a set of quantum gates that prepares a quantum state, capturing the salient features of the electron correlation. uCCSD allows us to leverage the strengths of quantum computing to solve complex chemical problems with a heuristic deeply rooted in the theory of classical quantum chemistry.

Despite its promise, the uCCSD ansatz encounters practical implementation challenges, especially when dealing with larger systems.\cite{Fedorov2022-ce, Tilly2022-pe} The number of entangling gates in uCCSD scales steeply with the number of qubits, which can quickly exceed the capabilities of current NISQ devices. 

As we navigate the NISQ era, the need for optimized circuit synthesis and compilation techniques is paramount. By allowing the execution of shallower circuits, these optimizations can potentially circumvent the limitations of current quantum hardware, striking a balance between performance and resource requirements.

Addressing these challenges, we introduce a suite of circuit optimization techniques designed for trapped-ion quantum devices. Our strategies employ optimal mapping from orbital to qubit, term reordering to minimize entangling gate counts, and the use of molecular point group symmetry to reduce redundant parameters. Additionally, we implement symmetry-based error mitigation post-selection methods, which substantially improve the quality of the quantum computation outcomes. Together, these approaches aim to significantly decrease circuit depth, broadening the scope of chemical systems within the reach of NISQ devices.

To demonstrate these techniques, we apply them to the highly symmetric benzene molecule. In this case, these methods facilitate the assembly of an 8-qubit circuit with nearly 70 entangling operations---a pushing the boundary of what is feasible with VQE circuits on quantum hardware. Importantly, these techniques not only extend our capabilities today but also open the door for the exploration of more complex chemical systems in the future.

\section{Methodology}
\subsection{Circuit Optimization Techniques}
In this section, we discuss the ansatz and circuit optimization techniques we use to reduce 
circuit depth. We follow closely with the approaches discussed in two previous studies\cite{Nam2020-ct, Zhao2023-um} on trapped-ion quantum computers. The uCCSD ansatz is written as 
\begin{equation}
    \label{eqn:uccsd_ansatz}
    \left|\Psi\right>=e^{T-T^\dagger}\left|\mathrm{HF}\right>
\end{equation}
in which $T$ is the cluster operator and $\left|\mathrm{HF}\right>$ is the Hartree-Fock state, which is a single bitstring encapsulating the mean-field solution that is both simple and efficient to prepare classically. The
cluster operator is
\begin{equation}
    \label{eqn:cluster_op}
    \begin{split}
        &T=T_1+T_2 \\
        &T_1=\sum_{ia}{t_i^aa^\dagger_aa_i} \\
        &T_2=\sum_{i<j,a<b}{t_{ij}^{ab}a^\dagger_ba_ja^\dagger_aa_i} \\
    \end{split}
\end{equation}
in which $a^\dagger_p$($a_p$) is the fermionic creation (annhilation) operator of the $p$-th orbital. 

Going from the analytical expression in Equation \ref{eqn:uccsd_ansatz} to an optimal quantum circuit,
the following steps need to be taken. First, orbitals are mapped to qubits. Second, excitation terms 
in the cluster operators are transformed to quantum gates one at a time. Third, circuit compilation
techniques are used to reduce the circuit depth. In this study, we use the Jordan-Wigner transformation\cite{Somma2002-au}
and write the fermonic operators to Pauli matrices. 
\begin{equation}
    \label{eqn:jwt}
    \begin{split}
        a^\dagger_p=Z_1\otimes\cdots\otimes Z_{p-1}\otimes (\frac{X_p-iY_p}{2})\otimes I_{p+1}\otimes\cdots\otimes I_n \\
        a_p=Z_1\otimes\cdots\otimes Z_{p-1}\otimes (\frac{X_p+iY_p}{2})\otimes I_{p+1}\otimes\cdots\otimes I_n \\
    \end{split}
\end{equation}
in which $X$, $Y$, $Z$ are the Pauli matrices, and $I$ is the identity matrix. 

Using JWT, a single excitation is written as
\begin{equation} 
    \label{eqn:jwt_single_excitation}
    \begin{split}
        a^\dagger_aa_i\rightarrow I_1\otimes\cdots\otimes I_{i-1}\otimes \left(\frac{X_i+iY_i}{2}\right)\otimes Z_{i+1}\otimes\cdots \\
        \otimes\cdots\otimes Z_{a-1}\otimes\left(\frac{X_a-iY_a}{2}\right)\otimes I_{a+1}\otimes\cdots\otimes I_n \\
    \end{split}
\end{equation}
in which we have assumed orbital index $i<a$. 

As one could see, the complexity of implementing the above depends on how close qubit $i$ and qubit $a$ are to each
other. If $i$ and $a$ are adjacent, then all the $Z$ matrices that go from index $i+1$ to $a-1$ disappear and this 
becomes a simple two-qubit gate, while if $i$ and $a$ are not adjacent to each other then the additional $Z$ matrices
make this term a multi-qubit gate, which increases the circuit depth. 

Since only particle-hole excitations are included in the uCCSD ansatz, we used a greedy search algorithm to 
map orbitals to qubits. In each iteration of the algorithm, it starts from a randomly selected excitation term from 
all the allowed excitations, then this algorithm searches for excitations that are most similar to the selected one in terms of orbital indices involved, 
and then maps these orbitals to qubits that are closest to each other. The search goes on until no similar 
excitations could be found, and then it goes to the next iteration with a new random selection. Such an algorithm 
ensures that interacting orbitals are mapped to qubits that are adjacent to each other, so that the number of 
Pauli-$Z$ matrices as in Equation \ref{eqn:jwt_single_excitation}, is minimized.

For each excitation, we need to implement the unitary operator $\prod_{\mu}{\mathrm{exp}[-i\theta P_{\mu}/2]}$, in which $P_{\mu} \in \{I, X, Y, Z\}^{\otimes 2N}$
is the Pauli word across all qubits, in which $N$ is the total number of spatial orbitals. For each double excitation, the product over
$\mu$ contains 8 terms, and we order them as $XXXY$, $XXYX$, $YXYY$, $YXXX$, $YYXY$, $YYYX$, $XYYY$, and $XYXX$. We choose such an order to allow for better circuit 
optimization. First, basis rotations are performed by applying $H$ or $S^\dagger H$ to
rotate qubits to the $X$ and $Y$ eigenbasis. Then the circuit shown below could be used to 
implement it, followed by the basis rotation back to the computational basis. The ordering
we choose has exactly 2 basis changes between adjacent terms, and it leaves only 2 $\mathrm{CNOT}$ gates between each term. 

\begin{center}
\label{circ:cx_ladder}
\begin{quantikz}
& \ctrl{3} & \qw  & \qw  & \qw  & \qw & \qw & \ctrl{3} & \qw \\
& \qw & \ctrl{2} & \qw & \qw  & \qw  & \ctrl{2} & \qw & \qw \\
& \qw  & \qw  & \ctrl{1} & \qw  & \ctrl{1} & \qw & \qw & \qw \\
& \targ{}  & \targ{}  & \targ{}  & \gate{R_z(\theta)} & \targ{} & \targ{} & \targ{} & \qw
\end{quantikz}
\end{center}

Finally, the resulting 2 $\mathrm{CNOT}$ gates between each term can be further optimized to 1 $\mathrm{CNOT}$ gate using the circuit identity below.  

\begin{center}
\label{circ:cx_ladder_opt}
\begin{quantikz}[row sep={7mm,between origins}]
& \ctrl{1} & \gate{H}  & \ctrl{1}  & \qw  \\
& \targ{} & \qw & \targ{} & \qw  \\
\end{quantikz}
\hspace{0.7em}=
\begin{quantikz}[row sep={7mm,between origins}]
& \gate{S} & \targ{} & \gate{S^\dagger}  & \gate{H}  & \qw  \\
& \gate{H} & \ctrl{-1} & \gate{S} & \gate{H} & \qw  \\
\end{quantikz}
\end{center}

\begin{figure}[htbp]
    \centering
    \includegraphics[width=0.5\linewidth]{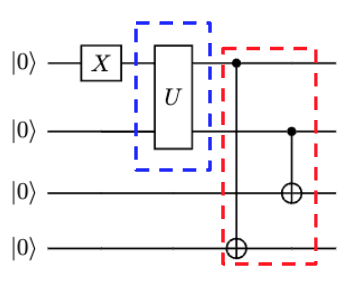}
    \caption{An example circuit for implementing paired excitations and transformation from spatial to
    spin orbitals. The $U$ operator in the blue dashed box implements a paired excitation between qubit
    0 and 1. The 2 $\mathrm{CNOT}$ gates in the red dashed box performs transformation from spatial to
    spin orbitals.}
    \label{fig:example_cirq}
\end{figure}

There are two types of double excitations in the uCCSD ansatz: paired and unpaired excitations. The former maintain electron pairs and the latter breaks them. As shown 
previously\cite{Vatan2004-jv}, each paired excitation can be implemented using only 2 $\mathrm{CNOT}$ gates, as 
they live in the space of spatial orbitals. Therefore, for a problem with $2N$ spin orbitals ($N$
spatial orbitals), we start from the $N$ qubits that all map to either $\alpha$ or $\beta$ spin orbitals, apply a $\mathrm{X}$ gate to each occupied orbitals, and implement all the paired excitations
for each occupied-virtual orbital pair. Then we apply $N$ $\mathrm{CNOT}$ gates targeting the rest
$N$ qubits. By doing so, the orbital occupation information is now shared across $\alpha$ and $\beta$
spin components. After this, all the unpaired excitations are implemented. An example circuit of 
4 qubits is shown in Figure \ref{fig:example_cirq}. 

\subsection{Utilizing Point Group Symmetry for Enhanced Computational Efficiency}

Point group symmetry is a potent tool in quantum chemistry, especially when applied to the unitary coupled cluster with singles and doubles (uCCSD)\cite{Stanton1991-sg,Cao2022-oe}. The classification of functions through molecular point group symmetry's irreducible representations (irreps) serves to identify integrals that are essentially zero, occurring when the product of their irreps does not encompass the totally symmetric representation. This determination is steered by the group multiplication table, indicating a null integral if the totally symmetric representation is missing in the irreps' product. This principle, while valid for one-electron integrals, is equally applicable to two-electron integrals, incorporating the irreps of the four involved functions.

\begin{figure}[htbp]
    \centering
    \includegraphics[width=\linewidth]{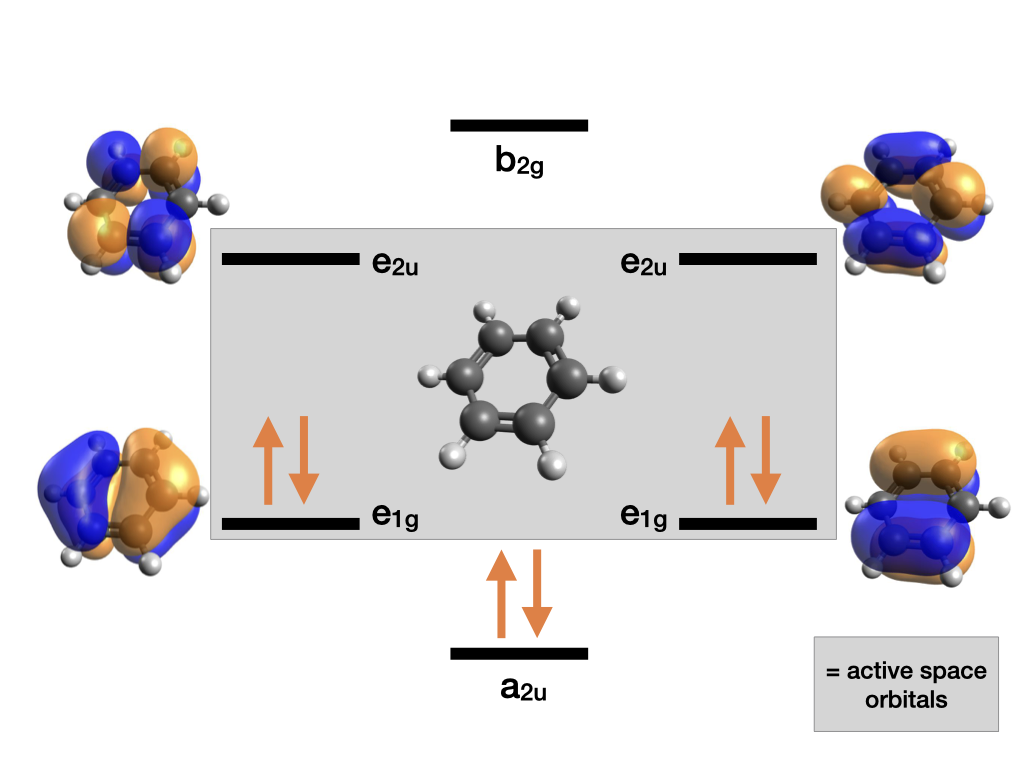}
    \caption{Representation of the benzene molecule's four $\pi$ orbitals. These orbitals form the active space for our quantum simulations, a choice driven by their major role in system energetics and chemical reactivity. Our decision to focus on these four orbitals (rather than a broader (6e, 6o) active space) was informed by balancing simulation accuracy with computational manageability, as indicated in Table \ref{tab:active_space_impact}.}
    \label{fig:orbitals}
\end{figure}

The benzene molecule, exhibiting $D_{6h}$ symmetry, serves as a case in point. The orbitals are illustrated in Figure \ref{fig:orbitals}. Only excitations in the totally symmetric representation (in this case $A_{1g}$) will result in non-zero components. For the (4e,4o) active space considered here, this eliminates, for example, the contributions from single excitation terms which are always of $E_{1g} \otimes E_{2u} = B_{1u}+B_{2u}+E_{1u}$ character -- clearly not in the totally symmmetric irrep! Note that to ease computation, we limit our focus to the Abelian subgroup $D_{2h}$, however, the above discussion holds.

\section{Results}

\subsection{Ansatze Selection}

Our ansatze selection process strove to strike an optimal balance between accuracy and circuit depth. We examined a range of uCC-based ansatze, using Qiskit's ideal statevector simulator and the QASM noise-free sampled simulator,\cite{Qiskit_contributors2023-vh} in scenarios both with and without point group symmetry. The impact of shot count was also evaluated, to understand its effect on accuracy and cost for potential hardware executions.

\begin{figure*}[htbp]
\centering
\includegraphics[width=\textwidth]{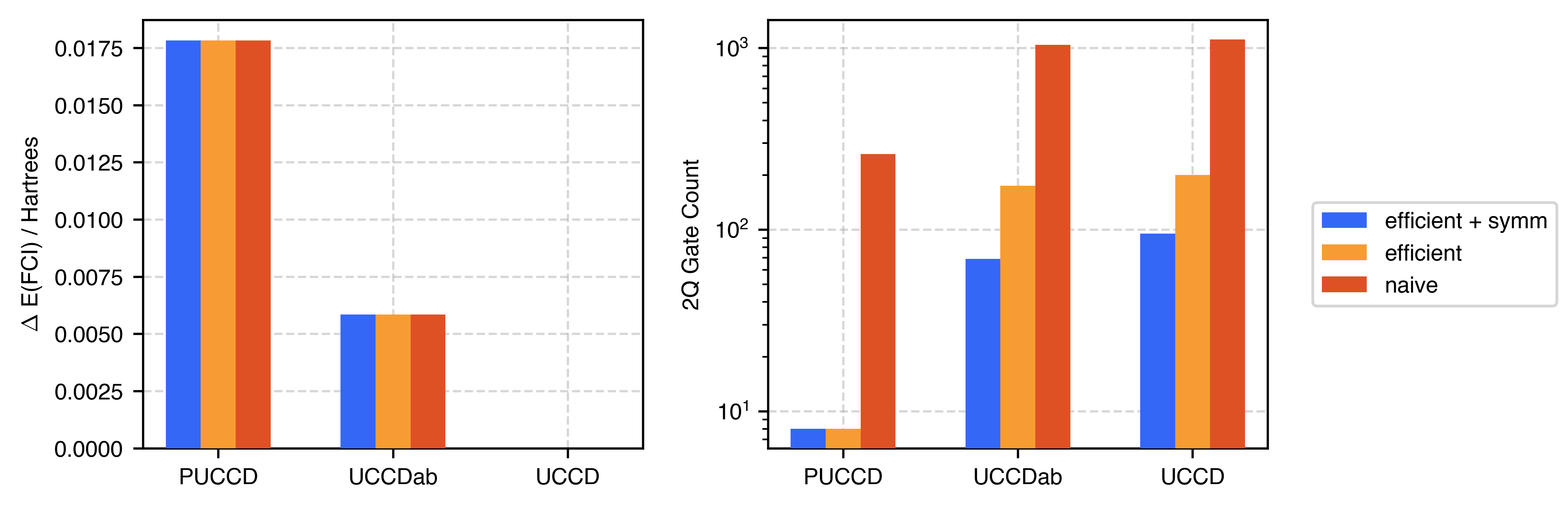}
\caption{Trade-off between accuracy (with respect to FCI) and 2-qubit gate depth for various ansatze. Our techniques (``efficient'' and ``efficient + symm'') maintain accuracy compared to baseline implementations, while significantly reducing circuit depth.}
\label{fig:ansatz_results}
\end{figure*}

In this study, four uCC-based ansatze---upCCD, uCCDab, uCCD, and uCCSD---played significant roles. The upCCD ansatz, limited to paired excitations, operates strictly within the seniority-zero subspace, simplifying computational demands while achieving acceptable accuracy.\cite{Lee2019-bu, Elfving2020-zx, Zhao2023-um} What we term the ``uCCDab'' ansatz, which includes only alpha-beta spin double excitations, avoids parallel spin excitations, yielding additional cost reduction with a minimal reduction in accuracy. uCCD embraces all double excitations and is found to be exact in this subspace. For that reason, the uCCSD ansatz, which considers all single and double excitations, is overspecified for this benzene model (and is also exact). While these restrictions in upCCD and uCCDab may limit their ability to depict some states, they offer a practical balance between circuit depth and accuracy. Importantly, all ansatze exclude single excitations—they do not influence the specific case of benzene—contributing to a reduction in circuit complexity. Among these ansatze, uCCD and uCCSD, being exact within the considered space (see Figure \ref{fig:ansatz_results}), have the potential to fully represent quantum states if sufficient computational resources are available. However, the inherent complexity of these ansatze makes them resource-intensive, and, in the case of uCCSD, somewhat redundant. The choice of ansatz, therefore, involves delicately balancing precision, circuit complexity, and resource efficiency.

\begin{figure}[htbp]
    \centering
    \includegraphics[width=\linewidth]{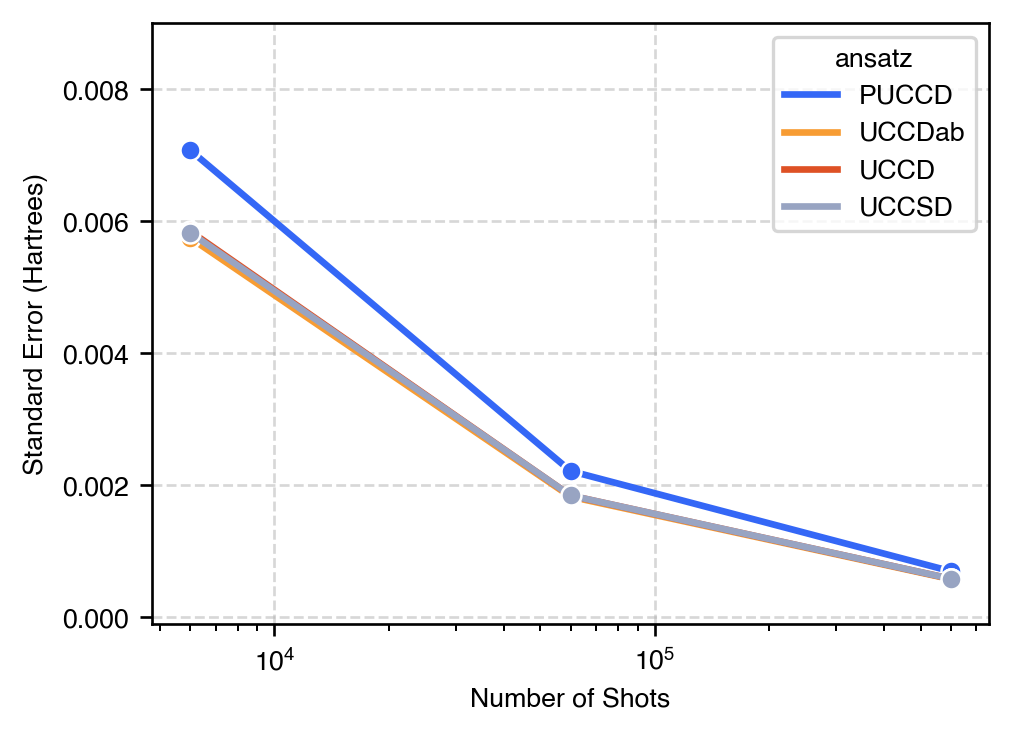}
    \caption{Energy standard error versus shot count for different ansatze. The uCCDab ansatz at 6000 shots presents an optimal trade-off, where statistical error and ansatz error are balanced ($\sim$ 6 mEh for both types of error). A 3x reduction in standard error would require a tenfold shot increase to 60000, indicating diminishing returns on computational resources.}
    \label{fig:std_err}
\end{figure}

The uCCDab ansatz emerged as the most suitable, yielding satisfactory accuracy without a significant increase in circuit depth. It is noteworthy that enforcing point group symmetry removed single excitations within the active space. Further, the shot counts were varied to balance between statistical reliability and computational resource management. These results of varying shot count and the energy standard error are given in Figure \ref{fig:std_err}.

The high symmetry of the benzene molecule ($D_{6h}$) led us to investigate the effects of applying point group symmetry in more depth. Operating within the Abelian $D_{2h}$ subgroup resulted in a substantial reduction in circuit depth without compromising accuracy. While extending the treatment of symmetry to non-Abelian point groups, such as the full $D_{6h}$ group, could yield additional savings, we did not explore this avenue in the present study.

The results of these explorations are summarized in Figure \ref{fig:ansatz_results}, highlighting the trade-offs involved in selecting different ansatze and their corresponding circuit implementation strategies.

\subsection{Active space orbitals}

The benzene molecule simulation's active space selection followed a careful balancing act, dictated by chemical intuition and pragmatic necessity. We decided to employ the molecule's nearest four frontier orbitals (Figure \ref{fig:orbitals}), given their primacy in many chemical reactions and because they are the largest component in the system energetics.\cite{Eriksen2020-ed} We explored additional active spaces in Table \ref{tab:active_space_impact}. Concretely, we used a (4e, 4o) active space focusing on the lowest lying $\pi$ orbitals. While using a (6e, 6o) active space may seem more appropriate, the exact CASCI energy difference between the two active spaces was 3 mEh, which is on the order of system noise. As the increased active space introduces additional noise through increased circuit depth and terms to measure, it did not seem efficient to consider the highest (lowest) lying frontier $\pi$ orbitals for this study. Focusing on the doubly degenerate orbitals also conveniently eliminated the need for single excitation components in our uCCSD ansatz as they vanish by symmetry.

This selection of active space deeply influenced the size and complexity of the quantum circuit we could construct. The impact of this selection can be seen in Table \ref{tab:active_space_impact}, revealing the trade-off between simulation accuracy and computational manageability.

\begin{table}[htbp]
\begin{threeparttable}
\caption{Comparison of exact energies (Hartrees) for several active space models in the minimal basis benzene molecule. The CAS(4,4) active space (gray) was selected as the best balance between accuracy and resource requirements, being around 5 mEh higher in energy from the largest active space considered which is on the same order as the standard error from the QPU runs. Circuit requirements are given for the efficient UCCDab ansatz with symmetry.}
\begin{tabular}{|R{1.3cm}|R{1.2cm}|R{0.7cm}|R{0.7cm}|R{0.8cm}|R{1.0cm}|}
\hline
\textbf{Active Space} & \textbf{Energy (Hartrees)} & \textbf{Qubits} & \textbf{2QGE}$^{\textrm{a}}$ & \textbf{QWC groups}$^{\textrm{b}}$ & \textbf{Var. Params}$^{\textrm{c}}$ \\ \hline
CAS(2,~2)  & -227.9059 & 4  & 4  & 5   & 1  \\ \hline
CAS(2,~4)  & -227.9103 & 8  & 10 & 45  & 3  \\ \hline
CAS(2,~6)  & -227.9103 & 12 & 16 & 112 & 5  \\ \hline
CAS(2,~8)  & -227.9104 & 16 & 56 & 282 & 9  \\ \hline
CAS(2,10) & -227.9105 & 20 & 94 & 699 & 13 \\ \hline
CAS(2,12) & -227.9106 & 24 & 232 & 1327 & 21 \\ \hline
\rowcolor{lightgray} CAS(4, 4)  & -227.9450 & 8  & 69 & 35  & 8  \\ \hline
CAS(4,~6)  & -227.9480 & 12 & 92 & 106 & 12 \\ \hline
CAS(4,~8)  & -227.9480 & 16 & 181 & 307 & 20 \\ \hline
CAS(4,10) & -227.9483 & 20 & 369 & 690 & 36 \\ \hline
CAS(4,12) & -227.9486 & 24 & 794 & 1468 & 60 \\ \hline
CAS(6,~6)  & -227.9480 & 12 & 87 & 106 & 13 \\ \hline
CAS(6,~8)  & -227.9486 & 16 & 325 & 331 & 31 \\ \hline
CAS(6,10) & -227.9505 & 20 & 844 & 700 & 67 \\ \hline
CAS(6,12) & -227.9518 & 24 & 1342 & 1372 & 95 \\ \hline
\end{tabular}
\label{tab:active_space_impact}
\begin{tablenotes}
\item[a] Number of two-qubit entangling gate equivalents.
\item[b] Number of qubit-wise commuting groups.
\item[c] Variational parameters in the ansatz.
\end{tablenotes}
\end{threeparttable}
\end{table}

\subsection{Results on IonQ Aria}

Our experimental implementation took place on IonQ's Aria Quantum Processing Unit (QPU). The Aria QPU employs trapped Ytterbium ions, with two states in the ground hyperfine manifold serving as qubit states. The states are modified by directing individual ions with 355 nm light pulses, thus instigating Raman transitions between the ground states that form the qubit. With the appropriate pulse configurations, we can realize both arbitrary single qubit gates and Mølmer-Sørenson type two-qubit gates.\cite{Sorensen2000-iu} The Aria QPU performance regarding fidelity and robustness has been documented elsewhere\cite{Lubinski2021-vq}.

The benzene molecule experiments were performed using IonQ's in-house quantum chemistry library, which is designed to facilitate the preparation and execution of variational quantum algorithms on trapped-ion QPUs. PySCF\cite{Sun2017-bw} was used to calculate molecular integrals and compute classical Full Configuration Interaction (FCI) reference energies. We optimized parameters classically using a statevector simulator from Qiskit,\cite{Qiskit_contributors2023-vh} bypassing iterative quantum-classical optimization loops. The circuits for benzene were assembled according to the previous discussion, and, with the classically optimized parameters, were run on IonQ Aria in order to evaluate the energy.

The results of the QPU and ideal energy evaluations can be seen in Table \ref{tab:results}. The ``raw'' energy results show an absolute error of 0.260 Hartrees. To mitigate some of this error, we employ two symmetry-based classical post-selection methodologies,\cite{Bonet-Monroig2018-xa} designated as ``particle'' and ``spin'' (see Table \ref{tab:results}). We perform these post-selection corrections on the computational basis terms (``Z-basis''), which does not introduce any additional quantum overhead. They operate entirely in the post-processing phase and are general to systems represented by Jordan-Wigner transformed (JWT) encoded fermions. 

The ``particle'' method assures the conservation of the total number of particles, or electrons, in the system by retaining outcomes where the number of ``1's'' in the bitstring corresponds to the actual electron count. The ``spin'' method focuses on spin symmetry conservation, ensuring that the number of particles in each spin manifold of the bitstring remains constant, hence prohibiting spin flips, which are forbidden by the ansatz. After post-selection, 4122 shots and 3101 shots of the original 6000 shots are retained for the ``particle'' and ``spin'' histograms in the computational basis. Despite this, we still observe a slight reduction in standard error in Table \ref{tab:results}, likely due to removal of unphysical high energy states.

The effectiveness of these classical post-selection strategies is evidenced in the significant reduction of error observed in the results. For instance, the ``spin'' post-selection method reduces the error from 0.260 Hartrees to 0.008 Hartrees, a level essentially at the QPU precision given the number of shots utilized (c.f. Figure \ref{fig:std_err}). This dramatic enhancement of accuracy underlines the power of these techniques in refining outcomes from quantum computations and advancing our understanding of the investigated molecular systems.

\begin{table*}[htbp]
\centering
\begin{tabular}{|l|l|l|c|}
\hline
Experiment & Post-selection & Total Energy (Hartrees) & Difference from Ideal (Hartrees) \\
\hline
Ideal / exact & None & -227.939117 & 0.000000 \\
Aria & None &  -227.679067 $\pm$ 0.004514 & 0.260050 \\
Aria & Particle &  -227.802476 $\pm$ 0.004065 & 0.136641 \\
Aria & Spin &  -227.930779 $\pm$ 0.001759 & 0.008338 \\
\hline
\end{tabular}
\caption{Comparison of total energies from ideal statevector simulations and real QPU experiments (IonQ Aria, 6000 shots), executed at the classically optimized parameters. Error is reported as plus or minus the standard error (SE). Note that reported SE represents the sampling error, not the total accuracy of the calculation as not all noise has been mitigated. Two classical post-selection methods, ``particle'' and ``spin'', have been employed to refine the QPU outcomes. These methods, operating on JWT-encoded fermionic systems in the computational basis, exploit the inherent symmetries of the system without introducing additional quantum overhead. The ``particle'' method enforces total particle number conservation, and the ``spin'' method ensures conservation of spin symmetry. Note that by enforcing spin symmetry we naturally enforce particle number symmetry as well. As observed, application of these post-selection methods can significantly reduce the error in energy measurements. Notably, enforcing spin symmetry reduces the error to 0.008 Hartrees, effectively achieving the QPU precision for the given number of shots.}
\label{tab:results}
\end{table*}

\section{Discussion}

Our work showcases a tangible advancement in the quest for practical quantum simulations of chemical systems, by employing novel circuit optimization techniques in trapped-ion quantum devices. Our strategies, including optimal mapping from orbital to qubit, term reordering, and the leveraging of molecular spin and point group symmetry, help mitigate the challenges that the unitary Coupled Cluster Singles and Doubles (uCCSD) ansatz encounters in real-world implementation, particularly when it comes to larger systems.

The significant reduction in circuit depth, as demonstrated by the benzene molecule simulation, suggests that a broader range of chemical systems can now be explored, expanding the scope and utility of current NISQ devices. It is pertinent to mention that our focus on benzene is primarily a choice of convenience, due to its high symmetry and consequent benefits for point group symmetry implementation. However, our techniques are readily extensible to other molecules, potentially even those with lower symmetry or larger size.

There remains a delicate balance to strike between accuracy and the practicalities of circuit implementation. As we venture into larger active spaces and more complex molecules, noise mitigation will play an increasingly crucial role until we approach the fault-tolerant era. While our techniques provide an efficient means of encoding and processing information, the ultimate success in the NISQ era will hinge on our ability to robustly mitigate noise.

\section{Conclusion}

Our study demonstrates the significant impact of optimal circuit synthesis and the potential benefits of all-to-all connectivity in trapped-ion quantum devices. By focusing on reducing the depth of uCCSD ansatz circuits, our strategies address one of the significant challenges plaguing the implementation of quantum simulations of chemical systems. The case study with the benzene molecule pushes the current boundaries of what is achievable with variational quantum eigensolver (VQE) circuits on quantum hardware.

At the same time, our results underline the importance of noise mitigation. Despite the promise of shorter and more efficient circuits, the realities of noisy quantum hardware demand resilient noise mitigation techniques to ensure accuracy. We show that simple approaches such as symmetry-based post-processing can yield promising results, but more robust strategies like the use of constraints on the reduced density matrices (RDMs)\cite{Rubin2018-yf} may be more generally applicable. Error mitigation is an ongoing area of research.\cite{Bonet-Monroig2018-xa,Montanaro2021-yv,Bennewitz2022-fo,Quek2022-ly,Kandala2019-rz,OBrien2022-aa,Weber2021-eg}

Ultimately, our work underscores the burgeoning potential of quantum simulations in the exploration of complex chemical systems. While we continue to face constraints from current hardware, the confluence of smart circuit design, noise mitigation, and quantum algorithm development are likely to pave the way for increasingly sophisticated quantum simulations.

\section{Data Availability}
The data presented in this manuscript are available from
the corresponding author upon reasonable request.

\section*{Acknowledgment} This work was jointly supported by Oak Ridge National Lab and IonQ, Inc. We are grateful to Torin Stetina for helpful comments, corrections, and discussion. The authors acknowledge DOE ASCR funding under the Quantum Computing Testbed Pathfinder program, FWP number ERKJ332.

\bibliographystyle{IEEEtran}
\bibliography{jjg_references}

\end{document}